\documentclass[oribibl,envcountsame]{llncs}

\usepackage{latexsym}

\def\qed{{\ \nolinebreak\hfill\mbox{$\Box$}}\smallskip}	

\newcommand{\SAT}{\mbox{\rm SAT}}
\newcommand{\QSAT}{\mbox{\rm QSAT}}

\newcommand{\oneinthree}{\mbox{\rm One-in-Three}}

\newcommand{\symor}{\mbox{\rm SymOR}_1}

\newcommand{\p}{\mbox{\rm P}}
\newcommand{\np}{\mbox{\rm NP}}

\begin{document}

\sloppy

\title{Dichotomy Theorems for Alternation-Bounded Quantified Boolean
Formulas\thanks{%
Research supported in part by grant NSF-CCR-0311021}}
\author{Edith Hemaspaandra}
\institute{
Department of Computer Science,\\
Rochester Institute of Technology,\\
Rochester, NY 14623, U.S.A.,\\
{\tt eh@cs.rit.edu}}

\maketitle
\thispagestyle{empty}

\begin{abstract}
In 1978, Schaefer proved his famous dichotomy theorem for generalized satisfiability
problems.  He defined an infinite number of propositional satisfiability problems,
showed that all these problems are either in P or NP-complete, and gave a simple
criterion to determine which of the two cases holds.  This result is surprising
in light of Ladner's theorem, which implies that there are an infinite number of
complexity classes between P and NP-complete (under the assumption that P is not
equal to NP).

Schaefer also stated a dichotomy theorem for quantified generalized Boolean
formulas, but this theorem was only recently proven by Creignou, Khanna, and Sudan,
and independently by Dalmau: Determining truth of quantified Boolean formulas is
either PSPACE-complete or in P.

This paper looks at alternation-bounded quantified generalized Boolean formulas.
In their unrestricted forms, these problems are the canonical problems complete
for the levels of the polynomial hierarchy.  In this paper, we prove dichotomy
theorems for alternation-bounded quantified generalized Boolean formulas, by showing
that these problems are either $\Sigma_i^p$-complete or in P, and we give a simple
criterion to determine which of the two cases holds.
This is the
first result that obtains dichotomy for an infinite number of
classes at once.

\noindent{\bf Keywords:}
quantified Boolean formulas,
computational complexity, 
Boolean constraints,
dichotomy,
polynomial hierarchy

\end{abstract}


\sloppy

\section{Introduction}
In 1978, Schaefer proved his famous dichotomy theorem for generalized
satisfiability
problems.  He defined an infinite number of propositional satisfiability
problems (nowadays often called Boolean constraint satisfaction
problems),
showed that all these problems are either in P or NP-complete,
and gave a simple
criterion to determine which of the two cases holds.  This result
is surprising
in light of Ladner's theorem, which implies that there are an infinite
number of
complexity classes between P and NP-complete (under the assumption that P is not equal to NP).

To make the discussion more concrete, we will quickly define
what a constraint is and what a constraint problem is.  Formal
definitions can be found in Section~\ref{s:prel}. In this paper,
we will  be talking about {\em Boolean} constraints.
See for example Feder and Vardi~\cite{fed-var:j:constraints} for a
discussion about general constraint satisfaction problems.

A {\em constraint} is a Boolean operator of fixed arity, specified
as a Boolean function.  For $C$ a constraint of arity $k$,
and $x_1, \ldots, x_k$ propositional variables (or constants),
$C(x_1, \ldots, x_k)$ is a {\em constraint application} of $C$.
For example, $\lambda xy. (x \vee y)$ is a constraint, and
$x_1 \vee x_2$ is a constraint application of this constraint.
Each finite set of constraints ${\cal C}$ gives rise
to a satisfiability problem $\SAT({\cal C})$: $\SAT({\cal C})$ is
the problem of, given a set of constraint applications of ${\cal C}$,
determining whether this set has a satisfying assignment.
We can view a set of constraint applications as a CNF formula.
For example, 2CNF-SAT corresponds to
$\SAT(\{\lambda xy . (x \vee y), \lambda xy . (x \vee \overline{y}),
\lambda xy . (\overline{x} \vee \overline{y})\})$.

Using constraint terminology, Schaefer's dichotomy theorem~\cite{sch:c:sat}
can now be formulated as follows: For any finite set of constraints
${\cal C}$, either $\SAT({\cal C})$ is in P, or
$\SAT({\cal C})$ is NP-complete.

In recent years, dichotomy theorems (or dichotomy-like theorems)
have been obtained for a number of other problems about logics.
For example, such theorems have be obtained for the problem
of determining whether a formula has exactly one
satisfying assignment~\cite{jub:c:usat}, the problem of
finding a satisfying assignment that satisfies a maximum number
of constraint applications~\cite{cre:j:max-sat}, the
problem of computing the number of satisfying
assignments~\cite{cre-heb:j:counting}, the problem of finding
the minimal satisfying assignment~\cite{kir-kol:j:minimal-sat},
the inverse satisfiability problem~\cite{kav-sid:j:inv-sat},
the equivalence problem~\cite{boe-hem-rei-vol:c:constraints},
the isomorphism problem~\cite{boe-hem-rei-vol:c:iso},
and  the complexity of propositional circumscription \cite{kir-kol:c:circum}. 
Khanna, Sudan, Trevisan, and Williamson examined the approximability of
some of these problems~\cite{kha-sud-tre-wil:j:approx-constraints}.
Consult the excellent monograph~\cite{cre-kha-sud:b:con} for an
almost completely up-to-date overview of dichotomy theorems for
Boolean constraint problems.

Schaefer also stated a dichotomy theorem for
quantified generalized formulas (or, equivalently, quantified sets
of constraint applications),  but this theorem was only recently proven by
Creignou, Khanna, and Sudan~\cite{cre-kha-sud:b:con}, and independently
by Dalmau~\cite{dal:t:dich}: Depending on the underlying finite set
of constraints, these problems are either PSPACE-complete or in P.

This paper looks at {\em alternation-bounded} quantified sets of constraint 
applications.  In their unrestricted forms, alternation-bounded
quantified Boolean formulas are the canonical problems
complete for the levels of the polynomial hierarchy. In this paper,
we prove dichotomy theorems for alternation-bounded quantified
sets of constraint applications, by showing that these problems
are either $\Sigma_i^p$-complete or in P,
and we give a simple criterion to determine which of the two cases
holds.  

The importance of these results is two-fold.
First of all,
unlike all previous results, our result obtains dichotomy for an
infinite number of classes
at once (namely, we prove dichotomy for each level of the polynomial
hierarchy).  Secondly,  Schaefer's dichotomy
theorem has proven very successful as a tool for proving NP-hardness.
After all, his theorem supplies us with
an infinite number of NP-complete variations of the already
often-used satisfiability problem. We expect that our dichotomy
theorems will likewise be  useful in proving problems hard
for higher levels of the polynomial hierarchy.  Though there
are not as many natural problems complete for higher
levels of the polynomial hierarchy as for NP, there are in
fact quite a few.  See the survey by M.~Schaefer and
Umans~\cite{sch-uma:j:ph}.

The rest of this paper is organized as follows. 
In Section~\ref{s:prel} we give the formal definitions of
constraints, constraint applications, complexity classes, 
and the various constraint problems that we
are interested in, and we will formally state Schaefer's dichotomy
theorem and the dichotomy theorem for quantified sets of 
constraint applications.
In Section~\ref{s:result} we will
prove the dichotomy theorems for  alternation-bounded quantified
constraint problems.

\section{Preliminaries}
\label{s:prel}

\subsection{Constraints}
We will use the terminology and notation
from~\cite{cre-kha-sud:b:con}.\footnote{It should be noted that
not all papers use this notation.  Many use the Schaefer notation
instead.   We have chosen to follow the~\cite{cre-kha-sud:b:con}
notation, because we will use some of their constructions.}

\begin{definition}
\begin{enumerate}
\item A {\em constraint} $C$ is a Boolean function from $\{0,1\}^k$ to
$\{0,1\}$, where $k > 0$. $k$ is the {\em arity} of $C$.
\item If $C$ is a constraint of arity $k$, and $z_1, z_2, \dots, z_k$
are (not necessarily distinct) variables, then $C(z_1, z_2, \dots,
z_k)$ is a {\em constraint application of $C$}.
\item If $C$ is a constraint of arity $k$, and for $1 \leq i \leq k$,
$z_i$ is a variable or a constant (0 or 1), then $C(z_1, z_2, \dots,
z_k)$ is a constraint application of $C$ {\em with constants}.
\end{enumerate}
When we want to be explicit about the variables
occurring in a set of constraint applications $S$, we 
will write $S(x_1, \ldots, x_n)$, to denote that the
variables of $S$ are in $\{x_1, \ldots, x_n\}$.
If we also want to be explicit about constants, we will
write $S(x_1, \ldots, x_n, 0, 1)$.
\end{definition}

Schaefer's generalized satisfiability problems can now be defined formally,
using constraint terminology.

\begin{definition}
Let ${\cal C}$ be a finite set of constraints.
\begin{enumerate}
\item $\SAT({\cal C})$ is the problem of deciding whether
a given set $S$ of constraint applications of ${\cal C}$
satisfiable, i.e., whether there exists an assignment to the variables
of $S$ that satisfies every constraint application in $S$.
\item $\SAT_c({\cal C})$ is the problem of deciding whether given
set $S$ of constraint applications of ${\cal C}$ with constants
is satisfiable.
\end{enumerate}
\end{definition}

As mentioned in the introduction, Schaefer proved that
all these problems are either in P or NP-complete.
It is also easy to determine which of these two cases hold.
This depends on simple properties of the constraints.

\begin{definition}
Let $C$ be a constraint.
\begin{itemize}
\item $C$ is {\em 0-valid} if $C(0, \ldots, 0) = 1$.
\item $C$ is {\em 1-valid} if $C(1, \ldots, 1) = 1$.
\item $C$ is {\em Horn} (or {\em weakly negative}) if $C$ is 
equivalent to a CNF formula where each clause has
at most one positive variable.
\item $C$ is {\em anti-Horn} (or {\em weakly positive}) if $C$ is 
equivalent to a CNF formula where each clause has
at most one negative variable.
\item $C$ is {\em bijunctive} if $C$ is equivalent to a 2CNF formula.
\item $C$ is {\em affine} if $C$ is equivalent to a XOR-CNF formula.
\item $C$ is {\em complementive} (or {\em C-closed}) if for every $s
\in \{0,1\}^k$, $C(s) = C(\overline{s})$, where $k$ is the arity of
$C$ and $\overline{s} = (1-s_1)(1-s_2) \cdots (1-s_k)$ for $s =
s_1s_2 \cdots s_k$.
\end{itemize}
Let $\cal C$ be a finite set of constraints.  We say ${\cal C}$ is
0-valid, 1-valid, Horn, anti-Horn, bijunctive, affine, or
complementive if {\em every} constraint $C\in{\cal C}$ is 0-valid,
1-valid, Horn, anti-Horn, bijunctive, affine, or complementive,
respectively.

\end{definition}

Schaefer's theorem can now be stated as follows.

\begin{theorem}[Schaefer~\cite{sch:c:sat}]
\label{t:SATdich}
Let ${\cal C}$ be a finite set of constraints.
\begin{enumerate}
\item If ${\cal C}$ is 0-valid, 1-valid, Horn, anti-Horn, affine, or 
bijunctive, then $\SAT({\cal
C})$ is in {\rm P}; otherwise, $\SAT({\cal C})$ is {\rm NP}-complete.
\item If ${\cal C}$ is Horn, anti-Horn, affine, or 
bijunctive, then $\SAT_c({\cal C})$ is in {\rm
P}; otherwise, $\SAT_c({\cal C})$ is {\rm NP}-complete.
\end{enumerate}
\end{theorem}

\subsection{Quantified constraint applications}

QBF is the problem of deciding whether a given  fully
quantified Boolean formula is true.
QBF is PSPACE-complete~\cite{sto-mey:c:poly}. 
This problem remains PSPACE-complete if we restrict
the Boolean formula to be in 3CNF~\cite{sto:j:poly}.
We use the following definition for quantified sets of
constraint applications.

\begin{definition}[\cite{cre-kha-sud:b:con}]
Let ${\cal C}$ be a finite set of constraints.
A {\em quantified ${\cal C}$ expression [with constants]} is an expression
of the form $Q_1 x_1 Q_2 x_2 \ldots Q_n x_n S(x_1, \ldots, x_n)$, where $S$
is a set of constraint applications of ${\cal C}$ [with constants],
and  $Q_i \in \{\exists, \forall\}$ for all $i$.
\end{definition}

We now define the constraint analogs of QBF.

\begin{definition}[\cite{cre-kha-sud:b:con}, Definition 3.9]
\begin{enumerate}
\item $\QSAT({\cal C})$ is the problem of deciding whether a given
quantified ${\cal C}$ expression
is true.
\item $\QSAT_c({\cal C})$ is the problem of deciding whether a
quantified ${\cal C}$ expression with constants
is true.
\end{enumerate}
\end{definition}

$\QSAT({\cal C})$ and $\QSAT_c({\cal C})$ exhibit dichotomy as well.
Remarkably, if $\SAT_c({\cal C})$ is in P, then so are
$\QSAT({\cal C})$ and $\QSAT_c({\cal C})$. In all other cases, 
$\SAT_c({\cal C})$ is NP-complete and
$\QSAT({\cal C})$ and $\QSAT_c({\cal C})$ are
PSPACE-complete.

\begin{theorem}[\cite{sch:c:sat,cre-kha-sud:b:con,dal:t:dich}]
\label{t:QSATdich}
Let ${\cal C}$ be a finite set of constraints.
If ${\cal C}$ is Schaefer, then $\QSAT({\cal C})$
and $\QSAT_c({\cal C})$ are in {\rm P};
otherwise, $\QSAT({\cal C})$ and
$\QSAT_c({\cal C})$ are {\rm PSPACE}-complete.
\end{theorem}

The history behind this theorem is rather interesting.
The dichotomy theorem for $\QSAT_{c}({\cal C})$ was stated without proof
by Schaefer~\cite{sch:c:sat}.  Schaefer mentioned that the proof relies
on the result that the set of true quantified 3CNF formulas
is PSPACE-complete.  Creignou et al.\ proved Theorem~\ref{t:QSATdich}
in~\cite{cre-kha-sud:b:con}.  The proofs of the PSPACE
lower bounds for $\QSAT_c({\cal C})$ are similar to
the NP-hardness proofs for $\SAT_c({\cal C})$.
It is shown by Creignou et al.~\cite{cre-kha-sud:b:con}, and independently
by Dalmau~\cite{dal:t:dich}, that $\QSAT_c({\cal C})$ polynomial-time
many-one reduces to  $\QSAT({\cal C})$.

\subsection{The Polynomial Hierarchy and Constraints}
\label{s:ph}

The polynomial-time hierarchy (polynomial hierarchy or PH for short)
was defined by
Meyer and Stockmeyer~\cite{mey-sto:c:reg-exp-needs-exp-space}. 

\begin{definition}[\cite{mey-sto:c:reg-exp-needs-exp-space}]
\begin{itemize}
\item $\Sigma_0^p = \Pi_0^p = \p$.
\item $\Sigma_{i+1}^p = \np^{\Sigma_{i}^p}$
\item $\Pi_{i+1}^p = {\rm co}\np^{\Sigma_{i}^p}$
\end{itemize}
\end{definition}

QSAT$_i$ is the set of all true fully quantified boolean formulas
with $i-1$ quantifier alternations, starting with an $\exists$ quantifier.
For all $i \geq 1$, QSAT$_i$ is complete for
$\Sigma^p_i$~\cite{sto-mey:c:poly}.
These problems remain $\Sigma_i^p$-complete if we restrict
the Boolean formula to be in 3CNF for $i$ odd and to 3DNF for $i$
even~\cite{wra:j:complete}.

To generalize $\QSAT_i$ to arbitrary sets of constraints,
it is important to realize that 3CNF formulas correspond
to sets of constraint applications, but 3DNF formulas do not.
Of course, a 3DNF formula is the negation of a 3CNF formula.
For $i$ even, we can view $\QSAT_i$ as the set
of all {\em false} fully quantified boolean formulas of the form
$\forall X_1 \exists X_2 \cdots \exists X_i \phi(X_1, \ldots, X_k)$,
where $X_1, \ldots, X_k$ are sets of variables.
Restricting $\phi$ to 3CNF in this view of $\QSAT_i$ will still
be $\Sigma^p_i$-complete.
We can now generalize $\QSAT_i$ to arbitrary sets of constraints.

\begin{definition}
Let ${\cal C}$ be a finite set of constraints.
\begin{itemize}
\item 
For all $i \geq 1$, a $\Sigma_i({\cal C})$ expression [with constants] is an
expression of the form
$\exists X_1 \forall X_2 \cdots Q_i X_i S(X_1, \ldots, X_i)$,
where $S$ is a set of constraint applications of
${\cal C}$ [with constants].  Here $X_1, X_2, ...$ are sets of variables. 
\item
For all $i \geq 1$, a $\Pi_i({\cal C})$ expression [with constants] is an
expression of the form
$\forall X_1 \exists X_2 \cdots Q_i X_i S(X_1, \ldots, X_i)$,
where $S$ is a set of constraint applications of ${\cal C}$ [with constants].
\end{itemize}
\end{definition}

\begin{definition}
Let ${\cal C}$ be a set of constraints.  Let $i \geq 1$.
\begin{enumerate}
\item For $i$ odd, $\QSAT_i({\cal C})$ is the problem of
deciding whether a given $\Sigma_i({\cal C})$ expression is true, and
$\QSAT_{i,c}({\cal C})$ is the problem of deciding whether a given
$\Sigma_i({\cal C})$ expression with constants is true.
\item For $i$ even, $\QSAT_i({\cal C})$ is the problem of
deciding whether a given
$\Pi_i({\cal C})$ expression is false, and
$\QSAT_{i,c}({\cal C})$ is the problem of deciding whether a given
$\Pi_i({\cal C})$ expression with constants is false.
\end{enumerate}
\end{definition}

\section{Dichotomy in the Polynomial Hierarchy}
\label{s:result}
The main proof technique for 
lower bounds on constraint problems is to
show that the problems
can simulate an already-known-to-be hard problem.
The freedom allowed in the simulations depends on the type of
problem considered.  For example, for satisfiability problems, we are
allowed to introduce existentially quantified auxiliary variables.
In~\cite{cre-kha-sud:b:con} terminology, this is known as
a ``perfect implementation.''

\begin{definition}[\cite{cre-kha-sud:b:con}]
\begin{enumerate}
\item
A set of constraint applications $S(X,Y)$ perfectly implements
constraint $C$ iff $C(X) \equiv \exists Y S(X,Y)$. 
\item 
A set of constraints ${\cal D}$ perfectly implements constraint
$C$ iff there exists a set of constraint applications of
${\cal D}$ that perfectly implements $C$.
\end{enumerate}
\end{definition}

Perfect implementations work well for satisfiability problems.

\begin{lemma}[\cite{cre-kha-sud:b:con}, 5.12, 5.16]
\label{l:CKSimplement}
\begin{enumerate}
\item If $\SAT({\cal C})$ is \np-hard and every constraint
in ${\cal C}$ can be perfectly implemented by ${\cal D}$, then
$\SAT({\cal D})$ is also \np-hard.
\item If $\QSAT({\cal C})$ is {\rm PSPACE}-hard and every constraint
in ${\cal C}$ can be perfectly implemented by ${\cal D}$, then
$\QSAT({\cal D})$ is also {\rm PSPACE}-hard.
\end{enumerate}
\end{lemma}

It is easy to see that the same construction works for
PH as well.

\begin{lemma}
\label{l:implement}
For all $i \geq 2$,
if $\QSAT_i({\cal C})$ is $\Sigma^p_i$-hard and every constraint
in ${\cal C}$ can be perfectly implemented by ${\cal D}$, then
$\QSAT_i({\cal D})$ is also $\Sigma^p_i$-hard.
\end{lemma}

\begin{proof}
Much like the corresponding proof of Lemma~\ref{l:CKSimplement} for
$\QSAT({\cal C})$.
Let $Q_1 X_1 Q_2 X_2 \cdots \exists X_i S(X_1, \ldots, X_i)$ be
a $\Sigma_i({\cal C})$ expression if $i$ is odd and
a $\Pi_i({\cal C})$ expression if $i$ is even.
For every constraint application  $A(Y) \in S$, replace $A(Y)$
by a set of constraint applications $U(Y,Z)$ of ${\cal D}$ such that
$A(Y) \equiv \exists Z U(Y,Z)$. Make sure that the $Z$ is
a set of new variables, and that all introduced sets of
new variables are disjoint.
Let $\widehat{S}$ be the resulting set of constraint applications
and let $\widehat{Z}$ be the set of all new variables.
Then $Q_1 X_1 Q_2 X_2 \cdots \exists X_i S(X_1, \ldots,  X_i)$ 
is true iff
$Q_1 X_1 Q_2 X_2 \cdots \exists X_i \exists
\widehat{Z} \widehat{S}(X_1, \ldots, X_i, \widehat{Z})$ is true.
\end{proof}

The dichotomy theorem for the case with constants now
follows much in the same way as in the case for general
quantified expressions.

\begin{theorem}
\label{th:dich-const}
Let ${\cal C}$ be a finite set of constraints and
let $i \geq 2$.
If ${\cal C}$ is Horn, anti-Horn, affine, or bijunctive,
then $\QSAT_{i,c}({\cal C})$ is in {\rm P};
otherwise, $\QSAT_{i,c}({\cal C})$ is $\Sigma^p_i$-complete.
\end{theorem}

\begin{proof}
The polynomial-time cases follow immediately from the fact
that if ${\cal C}$ is Horn, anti-Horn, affine,
or bijunctive, then even $\QSAT_{c}({\cal C})$ is in {\rm P} 
(Theorem~\ref{t:QSATdich}).
It is also immediate that $\QSAT_{i,c}({\cal C})$ is in $\Sigma^p_i$.

It remains to show the $\Sigma_i^p$ lower bounds.  
We closely follow the proof that $\QSAT({\cal C})$ is
PSPACE-hard from~\cite[Theorem 6.12]{cre-kha-sud:b:con}.

Recall from Section~\ref{s:ph} that the 3CNF version
of $\QSAT_i$ is complete for $\Sigma^p_i$.  In constraint
terminology, $\QSAT_i({\cal D})$ is
complete for $\Sigma^p_i$, where
${\cal D} = \{\lambda xyz. x \vee y \vee z,
\lambda xyz. x \vee y \vee \overline{z},
\lambda xyz. x \vee \overline{y} \vee \overline{z},
\lambda xyz. \overline{x} \vee \overline{y} \vee \overline{z}\}$.

In addition, the constraint $\oneinthree$ (which is defined as the ternary
Boolean function that is true if and only if exactly one
of its three arguments is true)
can perfectly implement any ternary
function~\cite{cre-kha-sud:b:con}.
Using Lemma~\ref{l:implement}, it follows that 
$\QSAT_{i}(\{\oneinthree\})$ is $\Sigma^p_i$-hard.
If ${\cal C}$ is not Horn, not anti-Horn, not affine, and not bijunctive,
then ${\cal C} \cup \{\lambda x.\overline{x},\lambda x.x\}$
perfectly implements $\oneinthree$~\cite{cre-kha-sud:b:con}.
It follows from Lemma~\ref{l:implement}
that $\QSAT_i({\cal C} \cup \{\lambda x. \overline{x},\lambda x.x\})$
is $\Sigma_i^p$-hard.  This implies that
$\QSAT_{i,c}({\cal C})$ is $\Sigma_i^p$-hard: Let
$Q_1 X_1 Q_2 X_2 \cdots \exists X_i S(X_1, \ldots, X_i)$ be
a quantified ${\cal C} \cup \{\lambda x. x, \lambda x. \overline{x}\}$
expression. If there exists a variable $x$ such that
both $x$ and $\overline{x}$ are in $S$,
then  $S \equiv 0$.  In that case, replace all of $S$ by 0.
Otherwise, for every variable $x$ such that $x \in S$
and $\overline{x} \not \in S$, replace this variable by 1,
and remove $x$ from $S$.
For every variable $x$ such that $\overline{x} \in S$ and
$x \not \in S$, replace $x$ by 0 and remove $\overline{x}$ from $S$.         
Call the resulting set of constraint applications $\widehat{S}$.
Then $\widehat{S}$ is a set of constraint applications of ${\cal C}$
with constants, and 
$Q_1 X_1 Q_2 X_2 \cdots \exists X_i S(X_1, \ldots, X_i)$  is true iff
$Q_1 X_1 Q_2 X_2 \cdots \exists X_i \widehat{S}(X_1, \ldots, X_i)$  is true.
\end{proof}

Far more effort is needed to prove the lower bounds for the case
without constants. Indeed, the remainder of this paper is dedicated 
to establishing this result.

\begin{theorem}
\label{th:dich-noconst}
Let ${\cal C}$ be a finite set of constraints and
let $i \geq 2$.
If ${\cal C}$ is Horn, anti-Horn, affine, or bijunctive,
then $\QSAT_{i}({\cal C})$ is in {\rm P};
otherwise, $\QSAT_{i}({\cal C})$ is $\Sigma^p_i$-complete.
\end{theorem}

\begin{proof}
The upper bounds follow from Theorem~\ref{th:dich-const}.
For the remainder of this proof, suppose that ${\cal C}$ is not Horn,
not anti-Horn, not affine,
and not bijunctive. We need to show that
$\QSAT_{i}({\cal C})$ is $\Sigma_i^p$-hard. 
Without loss of generality, we assume that no constraint in
${\cal C}$ is a constant function.  (Since such constraints
are bijunctive, we can simply remove them.)

We will prove that $\QSAT_{i}({\cal C})$ is $\Sigma_i^p$-hard
by a case distinction 
that depends on whether or not ${\cal C}$ is 0-valid, 
1-valid, and/or complementive.  In all cases, we will
reduce $\QSAT_{i,c}({\cal C})$ to $\QSAT_{i}({\cal C})$. 

\begin{description}
\item[${\cal C}$ is 0-valid and not complementive]
In this case, ${\cal C}$ perfectly implements
the constraint
$\lambda xy.\overline{x} \vee y$~\cite[Lemma 5.41]{cre-kha-sud:b:con}.

As a starting point, we will first review the reduction from
$\QSAT_c({\cal C})$ to $\QSAT({\cal C})$
from~\cite[Theorem 6.12]{cre-kha-sud:b:con} for the case that
${\cal C}$ is 0-valid and not complementive.
The main observation needed for this reduction is that
$\forall y \{\overline{f} \vee y,
\overline{y} \vee t\}$ is equivalent to $\overline{f} \wedge t$.

Let $Q_1 x_1 \cdots Q_n x_n S(x_1, \ldots, x_n, 0, 1)$ be a 
quantified ${\cal C}$ expression with constants. Using
the observation above, it is easy to see that this expression
is equivalent to the quantified ${\cal C} \cup  \{\lambda xy.
\overline{x} \vee y\}$
expression $\exists f \exists t \forall y Q_1 x_1 \cdots Q_n x_n$
$\left [ S(x_1, \ldots, x_n, f, t) \cup
\{\overline{f} \vee y, \overline{y} \vee t\} \right ]$.

Thus, $\QSAT({\cal C} \cup \{\lambda xy. \overline{x} \vee y\})$ 
is PSPACE-hard. 
Since ${\cal C}$ perfectly implements $\lambda xy. \overline{x} \vee y$,
it follows by Lemma~\ref{l:CKSimplement}
that $\QSAT({\cal C})$ is PSPACE-hard.

Note that this construction does {\em not} prove that
$\QSAT_i({\cal C})$ is $\Sigma_i^p$-complete, since 
the construction turns
a $\Sigma_i({\cal C})$ expression with constants into a
$\Sigma_{i+2}({\cal C})$ expression for $i$ odd,
and a $\Pi_i({\cal C})$ expression with constants into
a $\Sigma_{i+1}({\cal C})$ expression for $i$ even.

However, it is easy to see that we can place $\exists f \exists t
\forall y$ anywhere in the quantifier string,
as long as $\exists f$ and $\exists t$ precede
$\forall y$.
This implies that, as long as the original expression
contains existential quantifiers followed by universal quantifiers,
we obtain the required reduction.
Formally,  for $i > 2$, we reduce $\QSAT_{i,c}({\cal C})$ to
$\QSAT_{i}(\cal C)$, by  mapping
$Q_1 X_1 \cdots \exists X_{i-2} \forall X_{i-1}
\exists X_i S(X_1, \ldots, X_i, 0, 1)$ to
$Q_1 X_1 \cdots \exists X_{i-2} \exists f \exists t \forall y
\forall X_{i-1} \exists X_i
\left [ S(X_1, \ldots, X_i, f, t) \cup
\{\overline{f} \vee y,
\overline{y} \vee t\} \right ]$.

Since ${\cal C}$ is 0-valid, we know from Theorem~\ref{t:SATdich}
that, under the assumption that $\p \neq \np$,
$\QSAT_{i,c}({\cal C})$ is not reducible to
$\QSAT_{i}({\cal C})$ for $i = 1$.
It remains to handle the case that $i = 2$.

Let $\forall X_1 \exists X_2 S(X_1, X_2, 0, 1)$ be 
a $\Pi_2({\cal C})$ expression with constants.
We claim that this expression is equivalent to the
following $\Pi_2({\cal C} \cup \{\lambda xy. \overline{x} \vee y\})$ expression:
\[\forall X_1 \forall y \forall z \exists f \exists t \exists X_2
\left [ S(X_1, X_2, f, t) \cup \{\overline{f} \vee y, \overline{z} \vee t\} \right ].\]

For the proof, note that
\begin{eqnarray*}
\forall X_1 \forall y \forall z \exists f \exists t \exists X_2 & &
\left [ S(X_1, X_2, f, t) \cup \{\overline{f} \vee y, \overline{z} \vee t\} \right ]\\
\mbox{iff} & &\\
\forall X_1 \exists f \exists t \exists X_2 & &
\left [ S(X_1, X_2, f, t) \cup \{\overline{f} \vee 0, 1 \vee t\} \right ],\\
\forall X_1 \exists f \exists t \exists X_2 & &
\left [ S(X_1, X_2, f, t) \cup \{\overline{f} \vee 0,  0 \vee t\} \right ],\\
\forall X_1 \exists f \exists t \exists X_2 & &
\left [ S(X_1, X_2, f, t) \cup \{\overline{f} \vee 1, 1 \vee t\} \right ]\mbox{, and}\\
\forall X_1 \exists f \exists t \exists X_2 & &
\left [ S(X_1, X_2, f, t) \cup \{\overline{f} \vee 1, 0 \vee t\} \right ]\\
\mbox{iff}& &\\
\forall X_1 \exists f \exists t \exists X_2 & &
\left [ S(X_1, X_2, f, t) \cup \{\overline{f}, t\} \right ]\\
\mbox{iff}& &\\
\forall X_1 \exists X_2 & &
S(X_1, X_2, 0, 1)
\end{eqnarray*}

Note that this construction can be generalized to all $i \geq 2$, by
mapping 
$Q_1 X_1 \cdots \exists X_{i-2} \forall X_{i-1}
\exists X_i S(X_1, \ldots, X_i, 0, 1)$ to
$Q_1 X_1 \cdots \forall X_{i-1} \forall y \forall z \exists f \exists t
\exists X_i
\left [S(X_1, \ldots, X_i, f, t) \cup
\{\overline{f} \vee y, \overline{z} \vee t\}  \right ]$.

\item[${\cal C}$ is 1-valid and not complementive]
In this case, we could simply state that the proof is similar
to the proof of the case that ${\cal C}$ is 0-valid and not complementive.
But rather than making the reader work through the previous case
to see that this is actually true, we will prove a
theorem (Theorem~\ref{t:complement})
which relates satisfiability problems for sets of constraint
applications of ${\cal C}$ [with constants] to the
satisfiability problems where the set of constraints
is replaced by a type of ``complement.''  This theorem
immediately implies the current case and
will also be useful in the case that ${\cal C}$ is
complementive.  We start with some definitions.

\begin{definition}
\begin{enumerate}
\item
Let $C$ be a k-ary constraint. Define constraint
$C^c$ as follows.  For all $s \in \{0,1\}^k$,
$C^c(s) = C(\overline{s})$, where, as in the definition
of complementive, $\overline{s} = (1 - s_1)(1-s_2) \cdots (1-s_k)$ for $s =
s_1s_2 \cdots s_k$.  Note that $C$ is complementive iff $C = C^c$.
\item Let ${\cal C}$ be a finite set of constraints.  
Define the set of constraints
${\cal C}^c$  as ${\cal C}^c = \{C^c \ | \ C \in {\cal C}\}$.
\item For $S$ a set of constraint applications of ${\cal C}$ with constants,
define $S^c$ as $\{C^c(z_1, \ldots, z_k) \ | \ 
C(z_1, \ldots, z_k) \in S\}$, where each $z_i$ is a variable or a constant.
\end{enumerate}
\end{definition}

\begin{theorem}
\label{t:complement}
\begin{enumerate}
\item For all $i \geq 1$, 
$\QSAT_i({\cal C}) \equiv^p_m \QSAT_i({\cal C}^c)$ and
$\QSAT_{i,c}({\cal C}) \equiv^p_m \QSAT_{i,c}({\cal C}^c)$.
\item $\QSAT({\cal C}) \equiv^p_m \QSAT({\cal C}^c)$ and
$\QSAT_c({\cal C}) \equiv^p_m \QSAT_c({\cal C}^c)$.
\end{enumerate} 
\end{theorem}

This theorem follows immediately from the
following lemma.

\begin{lemma}
\label{l:complement}
Let ${\cal C}$ be a finite set of constraints and let
$Q_1x_1 \cdots Q_n x_n S(x_1, \ldots, x_n,0,1)$ be a quantified
${\cal C}$ expression with constants. Then
$Q_1x_1\cdots Q_n x_n S(x_1, \ldots, x_n,0,1)$ is true 
if and only if
$Q_1x_1 \cdots Q_n x_n S^c(x_1, \ldots, x_n,1,0)$ is true.
\end{lemma}

\begin{proof}
The proof is by induction on $n$, the number of variables
in $S$.  For $n = 0$,
by definition of $S^c$, $S(0,1)  = S^c(1,0)$.
Now let $n > 0$, and suppose the claim holds for $n-1$.

If $Q_1 = \forall$, then
$Q_1x_1 \cdots Q_nx_n S(x_1, \ldots, x_n,0,1)$ is true
if and only if both
$Q_{2} x_{2} \cdots Q_n x_n S(0, x_2, \ldots, x_{n}, 0,1)$ and 
$Q_{2} x_{2} \cdots Q_n x_n S(1, x_2, \ldots, x_{n}, 0,1)$ are
true.  By induction, this is the case if and only
if both
$Q_{2} x_{2} \cdots Q_n x_n S^c(1, x_2, \ldots, x_{n}, 1,0)$ and
$Q_{2} x_{2} \cdots Q_n x_n S^c(0, x_2, \ldots, x_{n}, 1,0)$ are true,
which holds if and only if 
$\forall x_1 \cdots Q_nx_n S^c(x_1, \ldots, x_n,1,0)$ is true.
The proof for $Q_1 = \exists$ is similar.
\end{proof}

\item[${\cal C}$ is 0-valid and complementive]

If ${\cal C}$ is complementive, $C^c = C$ for all $C \in {\cal C}$.
The following corollary follows immediately from 
Lemma~\ref{l:complement}.

\begin{corollary}
\label{c:complement}
Let ${\cal C}$ be complementive. 
Let $S(x_1, \ldots, x_n, 0, 1)$ be a set
of constraint applications of ${\cal C}$ with constants.  Then 
$Q_1x_1Q_2x_2 \cdots Q_nx_n S(x_1, \ldots, x_n, 0, 1)$ iff 
$Q_1x_1Q_2x_2 \cdots Q_nx_n S(x_1, \ldots, x_n, 1, 0)$.
\end{corollary}

From~\cite[Lemma 5.41]{cre-kha-sud:b:con}, we know that ${\cal C}$ perfectly 
implements the constraint $\symor$, which is defined as
the constraint $\lambda xyz. (\overline{x} \wedge (\overline{y} \vee z)) \vee
(x \wedge (\overline{z} \vee y))$.

Using Corollary~\ref{c:complement}, we can now construct a reduction that
is similar to the one described at the end of
the case that ${\cal C}$ is 0-valid and not complementive
to reduce $\QSAT_{i,c}({\cal C})$ to $\QSAT_i({\cal C})$.

Let $Q_1 X_1 \cdots \exists X_{i} S(X_1, \ldots, X_i, 0, 1)$
be a $\Sigma_i({\cal C})$ expression with constants if
$i$ is odd, and a $\Pi_i({\cal C})$ expression with constants if
$i$ is even.  We claim that this expression is true
if and only if the following expression is true:
\begin{eqnarray*}
\lefteqn{Q_1 X_1 \cdots \forall X_{i-1} \forall x \forall y \forall z
\exists f \exists t \exists X_i}\\
& & \left [S(X_1, \ldots, X_i, f, t) \cup
\{\symor(x, f, y), \symor(x, z, t)\} \right ].
\end{eqnarray*}

For the proof, note that
\begin{eqnarray*}
\lefteqn{Q_1 X_1 \cdots \forall X_{i-1} \forall x \forall y \forall z
\exists f \exists t \exists X_i}\\
& & \left [S(X_1, \ldots, X_i, f, t) \cup
\{\symor(x, f, y), \symor(x, z, t)\}  \right ]
\end{eqnarray*}
if and only if
$Q_1 X_1 \cdots \forall X_{i-1}  \forall y \forall z
\exists f \exists t \exists X_i
\left [S(X_1, \ldots, X_i, f, t) \cup
\{\overline{f} \vee y, \overline{z} \vee t\}  \right ]$  and 
$Q_1 X_1 \cdots \forall X_{i-1}  \forall y \forall z
\exists f \exists t \exists X_i
\left [S(X_1, \ldots, X_i, f, t) \cup
\{\overline{y} \vee f, \overline{t} \vee z\}  \right ]$.

As in the 0-valid and not complementive case, this holds if
and only if
\begin{eqnarray*}
Q_1 X_1 \cdots \forall X_{i-1} 
\exists f \exists t \exists X_i & & \left [S(X_1, \ldots, X_i, f, t) \cup
\{\overline{f}, t\}  \right ]  \mbox{ and }\\
Q_1 X_1 \cdots \forall X_{i-1} 
\exists f \exists t \exists X_i & &  \left [S(X_1, \ldots, X_i, f, t) \cup
\{\overline{t}, f\}  \right ]\\
\mbox{iff} & &\\
Q_1 X_1 \cdots \forall X_{i-1} 
\exists X_i & &  S(X_1, \ldots, X_i, 0, 1)  \mbox{ and }\\
Q_1 X_1 \cdots \forall X_{i-1} 
\exists X_i & &  S(X_1, \ldots, X_i, 1, 0)\\
\mbox{iff} & & \mbox{(by complementivity)} \\
Q_1 X_1 \cdots \forall X_{i-1} 
\exists X_i & &  S(X_1, \ldots, X_i, 0, 1) 
\end{eqnarray*}

\item[${\cal C}$ is not 0-valid, not 1-valid, and complementive]

In this case, ${\cal C}$ can perfectly implement
$\lambda xy. x \oplus y$~\cite[proof of Lemma~5.24]{cre-kha-sud:b:con}.

Using Corollary~\ref{c:complement},
it suffices to replace $0$ by $f$, $1$ by $t$ and to
add $\exists f \exists t \{f \oplus t\}$. 
However, this existential
quantification has to be added at the start of the expression.
For example, consider the (false) expression $\forall x \{x = 0\}$.
Adding the existential quantification at the end of
the quantifier string will
give $\forall x \exists f \exists t \{x = f, f \oplus t\}$,
which is true.
(It doesn't matter whether $t$ is set to 0 and $f$ to 1 or 
vice-versa (by complementivity). However, we need to always
look at the same assignment.)
Adding the existential quantifiers at the start of the expression
gives a reduction from $\QSAT_{i,c}({\cal C})$ to
$\QSAT_{i}({\cal C})$ for $i$ odd.  More formally, if
$i$ is odd, we map
\[\exists X_1 \forall X_2 \cdots \exists X_i S(X_1, \ldots, X_i, 0,1)\]
to
\[\exists f \exists t \exists X_1 \forall X_2 \cdots \exists X_i
 \left [ S(X_1, \ldots, X_i, f,t) \cup \{f \oplus t\}  \right ].\]
Note that this reduction also works for $i = 1$.

We will now show how to reduce $\QSAT_{i,c}({\cal C})$ to
$\QSAT_{i}({\cal C})$ for $i$ even.  Let
$S(X_1, \ldots, X_i, 0,1)$ be a set of constraint applications of ${\cal C}$
with constants.  We map 
\[\forall X_1 \exists X_2 \cdots \exists X_i S(X_1, \ldots, X_i, 0,1)\]
to
\[\forall b \forall X_1  \exists X_2  \cdots
\exists X_i  \exists b'  \left [ S(X_1, \ldots, X_i, b,b')  \cup
\{(b \oplus b')\}  \right ]
\]

That this is indeed a reduction follows immediately from
Corollary~\ref{c:complement}.

\item[${\cal C}$ is not 0-valid, not 1-valid, and not complementive]
Let $A \in {\cal C}$ be 
not 0-valid, $B \in {\cal C}$ be not 1-valid,  and
$C \in {\cal C}$ be not complementive.  Recall that we may assume
that all constraints in ${\cal C}$ are satisfiable.
Let $s_A$ be a satisfying assignment for $A$,
let $s_B$ be a satisfying assignment for $B$, and
let $s_C$ be a satisfying assignment for $C$ such that
$\overline{s_C}$ is not a satisfying assignment for $C$.
Let $\widehat{A}(x,y)$ be a constraint application of 
$A$ defined as follows:  $\widehat{A}(x,y) = A(z_1, \ldots, z_k)$, where
$z_i = x$ if $(s_A)_i  = 0$, and $z_i = y$ if $(s_A)_i = 1$.
Define $\widehat{B}(x,y)$ from $B$ and $s_B$ and $\widehat{C}(x,y)$ from
$C$ and $s_C$ in the same way.  Then
$\widehat{A}(0,0) = 0, \widehat{A}(0,1) = 1$,  
$\widehat{B}(0,1) = 1, \widehat{B}(1,1) = 0$,  
$\widehat{C}(0,1) = 1$, and $\widehat{C}(1,0) = 0$.
Now consider the set of constraint applications
$\{\widehat{A}(f,t), \widehat{B}(f,t), \widehat{C}(f,t)\}$.
It is easy to see that this set perfectly implements 
$\overline{f} \wedge t$.

For our reductions, we need to replace 0 by $f$, 1 by $t$, and
add $\exists f\exists t \{\overline{f} \wedge t\}$.
Note that, unlike the previous case, we can add the 
existential quantifiers anywhere in the quantifier string,
since  $\exists f\exists t \{\overline{f} \wedge t\}$ completely
fixes the truth assignment to $f$ and $t$.

More formally, to reduce $\QSAT_{i,c}({\cal C})$ to
$\QSAT_{i}({\cal C})$, map
$Q_1 X_1 Q_2 X_2 \cdots \forall Q_{i-1}\exists X_i
S(X_1, \ldots, X_i, 0, 1)$ to\\ $Q_1 X_1 Q_2
X_2 \cdots \forall Q_{i-1} \exists X_i \exists f \exists t  \left [
S(X_1, \ldots, X_i, f, t) \cup \{\overline{f} \wedge t\}  \right ]$.
This shows that
${\QSAT_i^p}({\cal C})$ is $\Sigma_i^p$-hard.

As in the previous case, this reduction will work for $i = 1$ as well.
Thus, our proofs also imply Schaefer's dichotomy theorem.
\end{description}
\end{proof}

\goodbreak

\end{document}